\documentclass[12pt]{article}
\usepackage{amssymb}
\usepackage{graphicx}
\parskip \baselineskip

\newcommand{\R}{\mathbb{R}}

\newcommand{\ol}{\overline}

\newcommand{\ra}{\rightarrow}
\newcommand{\Ra}{\Rightarrow}

\newcommand{\name}[2]{{#1}{\scriptsize{#2}}}

\begin{document}
\begin{center}
{\bf COMPACTLY GENERATING ALL SATISFYING TRUTH ASSIGNMENTS OF A HORN FORMULA}

\name{M}{ARCEL} \name{W}{ILD}
\end{center}

\begin{quote}
A{\scriptsize BSTRACT} : {\footnotesize As instance of an overarching principle of exclusion an algorithm is presented that compactly (thus not one by one)  generates all models of a Horn formula. The principle of exclusion can be adapted to generate only the models of weight $k$. We compare and contrast it with constraint programming, $0,1$ integer programming,  and binary decision diagrams.}
\end{quote}

{\bf 1. Introduction}

This introduction soon jumps into medias res with a concrete example of a Horn formula on six variables, and a $(4 \times 6)$-table that complactly represents all its models ($=$ satisfying truth value assignments). We then indicate applications that benefit from the possibility to efficiently produce or count all models $X$ of a Horn-formula; or alternatively only the $X$'s with $|X|=k$ for some prescribed integer $k$. Afterwards the section break-up displays the article's fine structure. Concepts only sloppily defined in the introduction will be formally defined in Section 2.

So consider this Horn formula $\varphi = \varphi (a_1, \cdots, a_6)$:
$$\varphi \quad : =\quad (\ol{a}_1 \vee \ol{a}_2 \vee \ol{a}_3 \vee a_5) \wedge (\ol{a}_1 \vee \ol{a}_2 \vee \ol{a}_3 \vee a_6) \wedge (\ol{a}_3 \vee \ol{a}_4 \vee \ol{a}_5 \vee a_6) \wedge (\ol{a}_1 \vee \ol{a}_3 \vee \ol{a}_6)$$
Setting $0: =$ {\tt False} and $1:=$ {\tt True} one verifies that say $t:= (a_1, a_2, a_3, a_4, a_5, a_6) = (1,0,1,1,0,0)$ is a model of $\varphi$, that is, $\varphi (t) =1$. The following table compactly represents {\it all} models of $\varphi$:

\hspace*{2cm} \begin{tabular}{|l|c|c|c|c|c|c|} \hline
$\rho_1 =$ & $2$ & $2$ & $0$ & $2$ & $2$ & $2$\\ \hline
$\rho_2=$ & $0$ & $2$ & $1$ & $n$ & $n$ & $2$ \\ \hline
$\rho_3 =$ & $1$ & $0$ & $1$ & $n$ & $n$ & $0$ \\ \hline
$\rho_4=$ & $0$ & $2$ & $1$ & $1$ & $1$ & $1$ \\ \hline \end{tabular}

Table 1

Each of the rows $\rho_i$ represents several models of $\varphi$. Namely, a label $2$ indicates that the corresponding entry is free to be $0$ or $1$, and the wildcard $nn$ means that at least one $0$ must be present there (thus $nn = 11$ is forbidden). For instance, the model $(1,0,1,1,0,0)$ from before belongs to $\rho_3$ (here $nn = 10$). Viewed as set systems the rows $\rho_i$ happen to be mutually disjoint, and so the number $N$ of models evaluates to
$$N = |\rho_1| + |\rho_2| + |\rho_3| + |\rho_4| = 32+4 \cdot 3 + 3 + 2 = 49$$
Also the number $N'$ of $4$-element models (say) is conveniently calculated as
$$N' = 10 +5+2+0 =17$$
{\it Generating} all models from Table 1 is just as easy but necessarily more time and space consuming.

As to applications, getting all models of a Horn formula comprises the following special cases:
\begin{enumerate}
\item[(i)] Get all sets of a closure system ${\cal C}$ from an implicational base $\Sigma$ of ${\cal C}$.
\item[(ii)] Get all sets of a simplicial complex ${\cal J}$ from a negative-clause base $\Theta$ of ${\cal J}$.
\end{enumerate}
Concerning (i), consider the closure system ${\cal C}$ of all subsemigroups of a semigroup $(W, \circ)$. Here an implicational base of ${\cal C}$ is  obtained as the family $\Sigma$ of all ``implications'' $\{a, b\} \ra \{a \circ b, b \circ a\}$ where $a,b$ range over $W$. Akin to general {\it implicational bases} that means that a subset $X \subseteq W$ is closed (i.e. a member of ${\cal C}$) if and only if it is $\Sigma$-{\it closed} in the sense that whenever a ``premise'' $\{a,b\}$ happens to be contained in $X$, then also the ``conclusion'' $\{a \circ b, b \circ  a\}$ must lie in $X$. Using the algorithm of Section 5, our closure system ${\cal C}$ can be compactly generated ``chunk-wise'' as in Table 1, rather than one-by-one. Not just semigroups, all finite universal algebras can be dealt with this way. As another example, Formal Concept Analysis [GW] is a data mining method that revolves around a closure system ${\cal C}$, called formal concept lattice, for which it is harder to find an imlicational base $\Sigma$. The original 1984 algorithm of Ganter produces ${\cal C}$ and $\Sigma$ simultaneously. It has been improved a number of times [KOV] but still is bound to the one-by-one generation of ${\cal C}$.

Concerning (ii), a set $\Theta$ of sets $A^\ast \subseteq W$ is a {\it negative-clause base} of a simplicial complex ${\cal J}$ on $W$, if for all $X \subseteq W$ it holds that : $X$ belongs to ${\cal J}$ if and only if $X \not\supseteq A^\ast$ for all $A^\ast \in \Theta$. An efficient way for calculating $|{\cal J}|$ from $\Theta$, or more specifically the face numbers $f_k: = |\{X \in {\cal J}: \ |X| = k \}|$, is useful in many situations, e.g. for getting the rank selection probabilities of a stack filter [W3].

Here comes the section break-up. Section 2 reviews basic material about Horn formulae and introduces some convenient set theoretic notation. In particular the $\varphi$ from before will be written as $\Sigma \cup \Theta$, where $\Sigma$ consists of the implications $\{1,2,3\} \ra \{5,6\}$ and $\{3,4,5\} \ra \{6\}$, and $\Theta$ consists of the set $\{1,3,6\}^\ast$. Section 3 explains in an informal way how Table 1 was derived from $\Sigma \cup \Theta$. The remaining sections more carefully distinguish between generating respectively counting models. Section 4 introduces the principle of exclusion (POE) which is a novel method for generating all models of a Boolean formula $\psi$ that comes as a conjunction of subformulas of suitable shape. It was already employed in [W1] but will be discussed in more coherent form here. In Section 5 the POE gets refined to the Horn $n$-algorithm which handles Horn formulae $\psi$. Section 6 aims at generating merely the $k$-element models of a Horn formula.
Section 7 resumes the discussion of the POE but now from the {\it counting} point of view. Akin to before this gets refined in Section 8 to counting all $k$-element models of a Horn formula. The final Section 9 positions the POE among other frameworks such as binary decision diagrams, branch and bound, constraint programming.

{\bf 2. On Horn formulae and closure systems}

A clause in propositional logic is called {\it Horn} if it contains at most one positive literal. We shall carefully distinguish the two ensuing subcases. Thus,  a formula of type

(NC) \qquad $\ol{a}_1 \vee \ol{a}_2 \vee \cdots \vee \ol{a}_n \qquad \quad (n\geq 1)$

respectively

(UI) \qquad $\ol{a}_1 \vee \ol{a}_2 \vee \cdots \vee \ol{a}_n \vee b \quad (\mbox{equivalently}: \ (a_1 \wedge a_2 \wedge \cdots  \wedge a_n) \ra b \ ) \qquad \quad (n\geq 0)$

will be called a {\it negative clause}, respectively {\it unit implication}. A {\it Horn formula} is any propositional formula that is  equivalent  to a conjunction of negative clauses and unit implications. For the time being we concentrate on (UI) and return to (NC) in a moment.

It is often convenient to combine unit implications with the same left hand side, for instance
$$((a \wedge c) \ra b)\quad  \wedge \quad ((a \wedge c) \ra d) \quad \wedge \quad ((a \wedge c) \ra e)$$ 
is equivalent to $(a \wedge c) \ra (b \wedge d \wedge e)$. If $t: \{a, b, c, d, e \} \ra \{0,1\}$ is any function, viewed as {\it truth value assignment} with $0=$ {\tt False} and $1=$ {\tt True}, then $t$ {\it satisfies} (or is a {\it model} of) $a \wedge c \ra b \wedge d \wedge e$ if and only if 
$$t(a) = 0 \quad \mbox{or} \quad t(c) =0 \quad \mbox{or} \quad t(a) = t(b) = t(c) = t(d)= t(e) =1$$
We shall mostly identify a truth value assignment $t$ with the subset $X$ of variables that have $t$-value $1$. For instance, if $W$ is our universe of variables, and $\{a, b, c, d, e\} \subseteq W$, then $X \subseteq W$ satisfies $\{a, c\} \ra \{b, d, e\}$ if and only if
$\{a, c\} \not\subseteq X \quad \mbox{or} \quad \{a, b, c, d, e \} \subseteq X$.
Generally, let $A = \{a_1, \cdots, a_n\}$ and $B = \{b_1, \cdots, b_m\}$ be subsets of $W$. We speak of

(Im) \qquad $A \ra B$ \quad (as formula: $(a_1 \wedge \cdots \wedge a_n) \ra (b_1 \wedge \cdots \wedge b_m) \ )$

as an {\it implication} with {\it premise} $A$ and {\it conclusion} $B$. Combining $m$ unit implications of type $n=0$ in (UI) yields $b_1 \wedge b_2 \wedge \cdots \wedge b_m$. This amounts to an implication with empty premise:

(Im$_0$) \qquad $\phi \ra \{b_1, b_2, \cdots, b_m\}$

Be aware that the  negative clauses $\ol{a}_1 \vee \cdots \vee \ol{a}_n$ in (NC) are not dual in any sense to the (unique if occuring) positive conjunction $b_1 \wedge b_2 \wedge \cdots \wedge b_m$ in (Im$_0$).
Let $\Sigma$ be any\footnote{Whereas there is information in $\phi \ra B_i$, there is none in $A_i \ra \phi$, and so these can be dropped. Furthermore, we henceforth silently assume that premise and conclusion are disjoint since $A \ra B$ is equivalent to $A \ra (B \setminus A)$.}  family of implications $A_i \ra B_i$.  One verifies at once that
$$\mbox{Mod} (\Sigma) \quad : = \quad \{ X \subseteq W | \ X \ \mbox{is a model of all members of}  \ \Sigma \}$$
is closed under intersections, i.e.

(1) \hspace*{4cm} $X \cap Y \in \ \mbox{Mod}(\Sigma) \ \mbox{for all} \ X, Y \in \ \mbox{Mod}(\Sigma).$

Moreover, $W$ satisfies all $A_i \ra B_i$ (since $B_i \subseteq W$), and so

(2) \hspace*{3cm} $W \in \mbox{Mod} (\Sigma)$, in particular $\Sigma$ is always satisfiable.

By (1) and (2),  $\mbox{Mod}(\Sigma)$ is a closure system and
$$cl (U)\quad : = \quad \bigcap \{X \in \mbox{Mod}(\Sigma) : U \subseteq X \}$$
is a {\it closure operator} on $2^W$. The {\it $\Sigma$-closure} $cl(U)$ can be computed [MR, Thm.10.3] in time $O(||\Sigma ||+w)$ where  $w:= |W|$ and $||\Sigma ||$ is defined as the sum of the cardinalities of all premises and conclusions of implications in $\Sigma$.
Let ${\cal C}$ be any closure system on a set $W$. Then a set $\Sigma$ of implications such that ${\cal C} = \mbox{Mod}(\Sigma)$ is called an {\it implicational base} of ${\cal C}$. We recommend [BM] as a survey which also offers a historical perspective including the frequent rediscovery of concepts. 

As to negative clauses $\ol{a}_1 \vee \ol{a}_2 \vee \cdots \vee \ol{a}_n$, in (NC) we choose the set notation $\{a_1, \cdots, a_n\}^\ast$. Thus, a truth assignment $t$ satisfies  $A^\ast:= \{a_1, \cdots, a_n\}^\ast$ if and only if $X:= t^{-1}(1) \subseteq W$ is a {\it noncover} for $A^\ast$ in the sense that $A^\ast \not\subseteq X$. 
For any set $\Theta$ of negative clauses $A^\ast$ it is clear that
$$\mbox{Mod}(\Theta): = \{X \subseteq W| \ X \ \mbox{is a noncover for all} \ A^\ast \in \Theta \}$$
is a simplicial complex, i.e. closed under subsets:

(3) \qquad $X \in \ \mbox{Mod}(\Theta)$ and $Y \subseteq X$, implies $Y \in \ \mbox{Mod}(\Theta)$.

Note that

(4) \qquad $\emptyset \in \ \mbox{Mod}(\Theta)$, in particular $\Theta$ is always satisfiable.

Let ${\cal J}$ be any simplicial complex on $W$. Then a set $\Theta$ of negative clauses $A^\ast$ will be called a {\it negative-clause base} of ${\cal J}$ if $\mbox{Mod}(\Theta ) = {\cal J}$.

We define a {\it Horn $h$-formula} as a $h$-element set $\Sigma \cup \Theta$ where $\Sigma$ consists of implications and $\Theta$ consists of negative clauses. As opposed to (2) and (4), $\Sigma \cup \Theta$ need not be satisfiable. Whether or not it is, can be settled with unit resolution; see [SS] for a concise account. An alternative method using {\it cl} from above is mentioned after the proof of Theorem 2.

{\bf 3. The Horn $n$-algorithm - first serving}

Here we get a first impression of the Horn $n$-algorithm by working through an ad hoc example. That lays the foundation for its detailed description and theoretic evaluation in Section 5.
Consider e.g. this Horn $3$-formula:
$$\begin{array}{lll}
\Sigma &= & \{ \ \{1,2, 3\} \ra \{5,6\}, \quad \{3,4,5\} \ra \{6\} \ \}\\
\\
\Theta &= & \{ \ \{1,3,6\}^\ast \ \} \end{array}$$
Generally let $W =\{1,2, \cdots, w\}$ (here $w = 6$) and $\mbox{Mod}_0 : = 2^W$. For $1 \leq i \leq h$ let $\mbox{Mod}_i$ be the family of all subsets $X \subseteq W$ that satisfy the first $i$ members of $\Sigma \cup \Theta$. In particular $\mbox{Mod}(\Sigma \cup \Theta) = \ \mbox{Mod}_h$. The main idea (to be elaborated in Section 4) is to calculate the subcollection $\mbox{Mod}_{i+1}$ from $\mbox{Mod}_i$ by discarding all $X \in \ \mbox{Mod}_i$ that falsify the $(i+1)$-th component. Any set $X \in \ \mbox{Mod}_i$ will be represented by its characteristic $0,1$-vector of length $w$, but whenever possible we use the label 2 to indicate that an entry is free to be 0 or 1. That is easy for $\mbox{Mod}_0$ which here is $r = (2,2,2,2,2,2)$. In order to represent $\mbox{Mod}_1$, let us split $r$ into the disjoint union of
$$r[n] = \{X \in r: \ \{1,2,3 \} \not\subseteq X\} \quad \mbox{and} \quad r[1] = \{X \in r: \ \{1,2,3\} \subseteq X\}.$$
We can compactly write these as
$$\begin{array}{lll}
r[n] & =& ( n,n,n,2,2,2)\\
r[1] & =& (1,1,1,2,2,2) \end{array}$$
with the understanding that the wildcard $nnn$ means ``at least one $0$''. Hence the letter $n$ which stands for {\it nul}. All $X \in r[n]$ trivially satisfy the implication $\{1,2,3\} \ra \{5,6\}$, but not all $X \in r[1]$ do. However, the good $X \in r[1]$ are easily pinned down and we get $\mbox{Mod}_1$ as the disjoint union of these ``$\{0,1,2,n\}$-valued rows:

\hspace*{2cm} \begin{tabular}{|c|c|c|c|c|c|} \hline
$n$ & $n$ & $n$ & $2$ & $2$ & $2$  \\ \hline
$1$ & $1$ & $1$ & $2$ & $1$ & $1$  \\ \hline \end{tabular}

Notice that all $X \in (1,1,1,2,1,1)$ satisfy the second implication $\{3,4,5\} \ra \{6\}$, but not all $X \in (n,n,n,2,2,2)$ do so. In order to pin down the  good $X$'s we split the row according to the third entry:
$$(n,n,n,2,2,2)\quad  =\quad (2,2, {\bf 0}, 2,2,2) \cup (n,n,{\bf 1}, 2,2,2).$$
All $X \in (2,2,0,2,2,2)$ satisfy $ \{3,4,5\} \ra \{6\}$, and those $X \in (n,n,1,2,2,2)$ that satisfy it are exactly the ones in $r_2 \cup r_3$:

\hspace*{2cm} \begin{tabular}{|l|c|c|c|c|c|c|} \hline
$r_1 =$ & $2$ & $2$ & $0$ & $2$ & $2$ & $2$\\ \hline
$r_2 =$ & $n_1$ & $n_1$ & $1$ & ${\bf n_2}$ & ${\bf n_2}$ & $2$ \\ \hline
$r_3=$ & $n_1$ & $n_1$ & $1$ & ${\bf 1}$ & ${\bf 1}$ & $1$\\ \hline
$r_4=$ & $1$ & $1$ & $1$ & $2$ & $1$ & $1$   \\ \hline \end{tabular}

From the above it is clear that $\mbox{Mod}_2 = r_1 \cup r_2 \cup r_3 \cup r_4$. 
All members of $r_1$ satisfy the negative clause  $\{1,3,6\}^\ast$, but no members of $r_4$ satisfy it. Hence $r_4$ needs to be cancelled. It is immediate that $\rho_4$ below comprises exactly the good $X\in r_3$, and not much harder to see that $\rho_2 \cup \rho_3$ comprises the good $X \in r_2$:

\hspace*{2cm} \begin{tabular}{|l|c|c|c|c|c|c|} \hline
$\rho_1 =$ & $2$ & $2$ & $0$ & $2$ & $2$ & $2$\\ \hline
$\rho_2=$ & ${\bf 0}$ & $2$ & $1$ & $n$ & $n$ & $2$ \\ \hline
$\rho_3 =$ & ${\bf 1}$ & $0$ & $1$ & $n$ & $n$ & $0$ \\ \hline
$\rho_4=$ & $0$ & $2$ & $1$ & $1$ & $1$ & $1$ \\ \hline \end{tabular}

Table 1

In summary,
$$\mbox{Mod}(\Sigma \cup \Theta ) = \ \mbox{Mod}_3 = \rho_1 \cup \rho_2 \cup \rho_3 \cup \rho_4$$
As seen in Section 1, all $X \in  \ \mbox{Mod}(\Sigma \cup \Theta )$ can be counted or generated from Table 1 in obvious ways. 
We close this section by giving the formal definition of a $\{0,1,2,n\}${\it -valued row} on a finite set $W$. It is a quadruplet
$$r\quad : =\quad (\mbox{zeros}(r), \ \mbox{ones}(r), \ \mbox{twos}(r), \ \mbox{nbubbles}(r))$$
such that
\begin{enumerate}
 \item [(5)] $W$ is the disjoint union of the sets zeros$(r)$, ones$(r)$, twos$(r)$, nbubbles$(r)$, where any one of these may be empty.
\item[(6)] If nbubbles$(r)$ is nonemtpy, then it is a disjoint union of $t \geq 1$ many sets $nb_1, \cdots nb_t$ (called $n$-bubbles) such that $\nu_i : = |nb_i| \geq 2$ for all $1 \leq i \leq t$.
\end{enumerate}
Thus $r$ can be visualized (up to permutation of the entries) as
\begin{enumerate}
\item[(7)] \quad  $r \quad = \quad \underbrace{(0, \cdots, 0}_{\alpha}, \underbrace{1, \cdots, 1}_{\beta}, \underbrace{2, \cdots, 2}_{\gamma}, \underbrace{n_1, \cdots, n_1}_{\nu_1}, \cdots, \underbrace{n_t, \cdots, n_t)}_{\nu_t}$
\end{enumerate}
By definition, $r$ {\it represents} the family of all sets $X \subseteq W$ satisfying
\begin{enumerate}
\item[(8)] \quad $X \cap \ \mbox{zeros}(r) = \phi$ and ones$(r) \subseteq X$ and $(\forall 1 \leq i \leq t) \ nb_i \not\subseteq X$.
\end{enumerate}
It will however be convenient to {\it identify} $r$ with the family of $X$'s satisfying (8). Then, clearly,
\begin{enumerate}
\item[(9)] \quad $|r|\quad =\quad 2^\gamma \cdot (2^{\nu_1} -1) \cdots (2^{\nu_t} -1)$.
\end{enumerate}

{\bf 4. The principle of exclusion aimed at generating}

A set $W$ with $|W|=w$ will be called a $w$-{\it set}. Formally, a {\it constraint} on a fixed $w$-set $W$ is a family ${\cal P} \subseteq 2^W$.  We say that $X \subseteq W$ {\it satisfies} the constraint ${\cal P}$ iff $X \in {\cal P}$. Equivalently a constraint can be defined as a Boolean function $b: \{0,1\}^W \ra \{0,1\}$ in that $X \subseteq W$ satisfies ${\cal P}$ if and only if its characteristic vector $x$ satisfies $b(x) =1$.
Proceeding with Boolean function terminology our task below could be defined as a specific constraint satisfaction problem (CSP). We touch upon the standard CSP line of attack in Section 9 but here we try another approach for which the set theoretic frameword is more convenient.

The task to find all $N$ {\it models} $X$ satisfying $h$ given constraints ${\cal P}_i$ amounts to determine ${\cal P}_1 \cap {\cal P}_2 \cap \cdots \cap {\cal P}_h$. For instance, ${\cal P}_i$ may be the constraint of being closed with respect to some implication $A_i \ra B_i$. Starting with the powerset $\mbox{Mod}_0 : = 2^W$ the {\it principle of exclusion}\footnote{Of course this has got nothing to do with Pauli's famous ``principle of exclusion'' known from physics. The name arose as a contrast to the well known principle of inclusion-exclusion.} ($POE$) generates
$$\mbox{Mod}_{i+1}\quad : =\quad \{X \in \ \mbox{Mod}_i | \ X \mbox{satisfies} \ {\cal P}_{i+1} \}$$
by {\it excluding} all duds $X$ (i.e. violating ${\cal P}_{i+1}$) from the family $\mbox{Mod}_i$ of ``partial models'' (i.e. satisfying  the first $i$ constraints).
 At the end $\mbox{Mod}_h$ equals ${\cal P}_1 \cap \cdots \cap {\cal P}_h$.  All of that is only efficient when $\mbox{Mod}_i$ can be compactly represented as union of disjoint ({\it multivalued}) {\it rows}  $r$. In the worst case $r$ is just a $0,1$-vector but usually $r$ comprises other symbols as well. For instance, in section 3 multivalued meant $\{0,1,2,n\}$-valued. A row of $\mbox{Mod}_h$ is called a {\it final} row.

The row collection $\mbox{Mod}_{i+1}$ arises from $\mbox{Mod}_i$ by {\it imposing} constraint ${\cal P}_{i+1}$ on each row $r \in \ \mbox{Mod}_i$. Imposing ${\cal P}_{i+1}$ on $r$ happens in exactly one of three ways:
\begin{enumerate}
 \item [(10)] If no $X \in r$ satisfies ${\cal P}_{i+1}$, then $r$ is deleted.
\item[(11)] Otherwise $r$ can sometimes be promoted to another row $r'$ which comprises exactly those $X \in r$ that satisfy ${\cal P}_{i+1}$. We call $r'$ a {\it trivial son} of $r$. In particular, if {\it all} $X \in r$ happen to satisfy ${\cal P}_{i+1}$ already, then $r' =r$.
\item[(12)] If $\{X \in r | \ X$ satisfies ${\cal P}_{i+1} \} \subseteq r$ cannot be modelled by a trivial son $r'$, one proceeds as follows:
 \begin{enumerate}
 \item [(12.1)] Row $r$ is {\it split} into disjoint {\it candidate sons} $r_j \ (1\leq j \leq s)$, i.e. each $X \in r$ is contained in exactly one $r_j$. Here $2 \leq s \leq cw$.
\item[(12.2)] If $r_j$ contains no member satisfying ${\cal P}_{i+1}$, then $r_j$ is deleted. Otherwise $r_j$ is altered (shrunk) and promoted to a {\it proper son} $r_j'$ that contains exactly those $X \in r_j$ that satisfy ${\cal P}_{i+1}$.
\end{enumerate}
\end{enumerate}

The $c$ in (12.1) is a global constant, i.e. depends only on the type of POE-application. For later purpose we define $s_{\max}$ as the maximum $s$ occuring in any fixed concrete run of POE. Often $cw$ an be substituted by $c$, for instance $c=3$ in the semigroup application of Section 1. If in (12.2) {\it all} candidate sons get killed, that amounts to (10). In a good use of the principle of exclusion the deletable rows, if any, should be recognized as soon as possible in order not to waste time on doomed successor rows.
To simplify later proofs it is convenient to postulate the following condition which so far always held anyway:
\begin{enumerate}
\item[(13)] The imposition of any constraint ${\cal P}_i$ upon any multivalued row $r$ of length $w$ costs $O(w^2)$.
\end{enumerate}

{\bf 4.1 Time assessment}

 A multivalued row $r$ is called {\it feasible} if it contains at least one model. In other words, $r \cap {\cal P}_1 \cap \cdots \cap {\cal P}_h \neq \emptyset$. The fact that a feasible row never gets killed will be referred to as the {\it consistency} of the principle of exclusion. We say that a particular installment of the principle of exclusion {\it avoids the deletion of rows} if (10) above never occurs.  

One of the benefits of the principle of exclusion is that for any integer $k \leq w$ and any row $r$ the number

(14) \qquad $Card(r,k) \quad : =\quad | \{ X \in r: \ |X| = k\}|$

of $k$-element members of $r$ is often easier to calculate than in other computing frameworks (Section 9). By focusing on $|X| \leq k$ respectively $|X| \geq k$ we similarly define $Card (r, \leq k)$ and $Card (r, \geq k)$. Thus Card$(r, \leq w)$ is just $|r|$ which, as in (9), was particularly easy to compute in all instances of POE so far. 

We say that a function $f(h,w)$ is {\it at least linear} in $w$ if for some constant $\ol{c}> 0$ it holds that $f(h,w) \geq \ol{c}w$ for all positive reals $h$ and $w$.

\begin{tabular}{|l|} \hline \\
{\bf Theorem 1:} Let $W$ be a $w$-set and let ${\cal P}_i \subseteq 2^W$ be constraints $(1 \leq i \leq h)$.  Fix\\
$k \in \{1,2, \cdots, w\}$. Suppose some ``old'' version of the principle of exclusion can be \\
employed  to produce disjoint multivalued rows whose union is the set of all models. \\
Further assume that for some function $f(h, w)$ which is at least linear in $w$, \\
it holds that:\\ \\
(a) For each row $r$ it costs time $O(f(h,w))$ to decide whether there is a model\\
\hspace*{.5cm} $X \in r$ with $|X| \leq k$.\\ \\ 
(b) If $r$ is a final row, then it costs $O(Card (r, \leq k) wf(h,w))$ to write down\\
\hspace*{.5cm} (in ordinary set notation) the sets $X \in r$ with $|X| \leq k$.\\ \\

Then the old version can be adjusted to a new one that avoids deleting rows and   \\
generates the  $N$ models $X \subseteq W$ with $|X| \leq k$ in time $O(f(h,w) + Nhwf(h,w))$. \\ \\ \hline 
 \end{tabular}

The requirement that all sets must have cardinality $\leq k$ cannot be treated as some extra constraint ${\cal P}_{h+1}$, because it will not be ``imposed'' the same way as the others. However, it is convenient to call $r$ {\it extra feasible} if there is a model $X \in r$ with $|X| \leq k$. For the special case $k = w$ in Theorem 1 ``extra feasible'' amounts to ``feasible''.

{\it Proof of Theorem 1.} The first row always is $(2,2,\cdots, 2)$, i.e. the powerset of $W$. If it is not extra feasible, this can by (a) be detected in time $O(f(h,w))$ and then there are $N=0$ models. That's the only reason why $O(f(h,w)+Nhwf(h,w))$ in Theorem 1 cannot be replaced by $O(Nhwf(h,w))$. An analogous argument in forthcoming theorems will not be repeated.

So assume that $(2,2,\cdots, 2)$ is extra feasible. We shall argue by  induction that the old algorithm can be renewed to make all promoted rows extra feasible, and so by consistency no promoted row can ever be deleted (having caused, together with its forfathers, much useless work).
 Hence, if $R$ is the number of final rows produced by the new algorithm, then the number of occurred impositions is at most $Rh$ (distinct finalized rows possibly having some of their $h$ forfathers coinciding). Below we shall show the ``core claim'' namely that the imposition of a constraint ${\cal P}_i$ upon a row still costs $O(wf(h,w))$ with the new algorithm. 
 
The {\it cost of all impositions} is thus $O(Rhwf(h,w))$. Because the sum of all $R$ numbers $Card (r, \leq k)$ when $r$ ranges over the (disjoint!) final rows is $N$, it follows from (b) that the {\it cost of generating} all  $(\leq k)$-element models from the final rows is $O(Nwf(h,w))$. In view  of $R \leq N$  adding up the two costs yields $O(Rhwf(h,w)) + O(Nwf(h,w)) = O(Nhwf(h,w))$ as claimed.

As to the core claim, let $r$ be extra feasible with say $X_0 \in r$ satisfying all constraints ${\cal P}_i \ (1 \leq i \leq h)$ and having $|X_0| \leq k$.  By (13) it costs $O(w^2)$ to impose a constraint ${\cal P}_i$ upon $r$. If $r$ gives rise to a trivial son $r'$, then by consistency still $X_0 \in r'$, and so $r'$ remains extra feasible. Suppose $r$ gives rise to the candidate sons $r_1, \cdots, r_s \ (s \geq 2)$. One of them, say $r_1$, must contain $X_0$.
Say $r_1, \cdots, r_t $ are exactly the extra feasible candidate sons. Even the old version of the principle of exclusion by consistency would promote $r_1, \cdots, r_t$ to proper sons $r'_1, \cdots, r'_t$. The new version additionally ensures, by testing all $r_j \ (1 \leq j \leq s)$ for extra feasibility, that {\it none} of $r_{t+1}, \cdots, r_s$ gets promoted.
By (a)  and since $s \leq cw$ by $(12.1)$, all of that costs $O(w^2) +sO(f(h,w)) = O(w^2)+ O(wf(h,w))$. Because $f(h,w)$ is at least linear in $w$ that reduces to $O(wf(h,w))$.
 \hfill $\blacksquare$

As to how the intermediate or final rows can be stored economically, see Section 7.1.

An analogous argument shows that Theorem 1 also holds when $\leq k$ is switched to $\geq k$, respectively $=k$, throughout.
Of course ``old $=$ new'' is possible in Theorem 1; then simply {\it one} algorithm that avoids deletion of rows is assessed.  

Call a multivalued row $r$ {\it weakly feasible} if for all $1 \leq i \leq h$ there is some $X_i \in r$ that satisfies ${\cal P}_i$. Thus ``feasible'' amounts to say that all $X_i$  $(1 \leq i \leq h)$ can be chosen {\it identical}. Because not all variants of the principle of exclusion allow a fast feasibility check, weak feasibility serves as a substitute: discarding rows which are not weakly feasible puts a lid on the number of  deletions. All the theorems to come concern only the feasibility of rows, but weak feasiblity will feature in our informal Section 9.

{\bf 5. The Horn $n$-algorithm - second serving}

Let us continue on a more technical level the discussion of the Horn $n$-algorithm begun in Section 3, making use of the POE framework displayed in Section 4. We first discuss the various cases that arise when an implication or negative clause is to be imposed on a row $r$. Afterwards Theorems 1 will be applied to the Horn $n$-algorithm.

So let $A \ra B$ be an implication, where $A$ (the premise) and $B$ (the conclusion) are w.l.o.g. nonvoid disjoint subsets of $W =\{1,2, \cdots, w\}$. It is to be imposed on a $\{0,1,2,n\}$-valued row $r$ indexed by $W$, as visualized in  (7).

{\bf Case 1:} $A \cap \ \mbox{zeros}(r) \neq \emptyset$, or $A$ wholy contains a $n$-bubble, or $B \subseteq  \mbox{ones}(r)$. In this case either all $X \in r$ have $A \not\subseteq X$, or all $X \in r$ have $B \subseteq X$. Whatever takes place, all $X \in r$ satisfy $A \ra B$, and so row $r$ carries over unaltered. Here are three instances of rows $r$ that all satisfy $\{1,2,3\} \ra \{5,7\}$. They correspond to the three mentioned subcases:

\hspace*{2cm} \begin{tabular}{|c|c|c|c|c|c|c|} 
$1$ & $2$ & $3$ &$4$ & $5$ & $6$ & $7$ \\ \hline
$1$ & $2$ & $0$ & $2$ & $2$ & $n$ & $n$\\ \hline
$n_1$ & $n_1$ & $n_2$ & $n_2$ & $n_2$ & $1$ & $0$\\ \hline
$n$ & $n$ & $1$ & $n$ & $1$ & $n$ & $1$ \\ \hline \end{tabular}

{\bf Case 2:} $A \subseteq \ \mbox{ones}(r)$ and ($B \cap \ \mbox{zeros}(r) \neq \emptyset$ or $B$ wholy contains a $n$-bubble). Then clearly $r$ needs to be cancelled.

{\bf Case 3:} $A \subseteq \ \mbox{ones}(r)$ and $B \cap \ \mbox{zeros}(r) = \emptyset$ and $B$ contains no $n$-bubble. Then we can switch all entries contained in $B$ to $1$ (while adjusting some others). Using the terminology of section 4 we thus obtain a trivial son $r' \subseteq r$ that satisfies $A \ra B$. For instance, for $\{1, 9\} \ra \{3,4,5,6\}$ we get:

\hspace*{.5cm} \begin{tabular}{c|c|c|c|c|c|c|c|c|c|}
& $1$ & $2$ & $3$ & $4$ & $5$ & $6$ & $7$ & $8$ & $9$  \\ \hline 
$r=$ & $1$ & $n_1$ & $n_1$ & $1$ & $2$ & $n_2$ & $n_2$ & $n_2$ & $1$ \\ \hline \end{tabular} \quad $\Ra$ \quad \begin{tabular}{c|c|c|c|c|c|c|c|c|c|} 
& $1$ & $2$ &$3$ & $4$ & $5$ & $6$ & $7$ & $8$ &$9$\\ \hline
$r'=$ & $1$ & $0$ & $1$ & $1$ & $1$ & $1$ & $n$ & $n$ & $1$ \\ \hline \end{tabular}

Given that $A \subseteq \ \mbox{ones}(r)$, either Case 2 or Case 3 takes place. Hence in view of Case 1 this is the only remaining possibility:

{\bf Case 4:} $A \not\subseteq \ \mbox{ones}(r)$, and $A \cap \ \mbox{zeros}(r) = \emptyset$, and $A$ does not wholy contain a $n$-bubble, and $B \not\subseteq \ \mbox{ones}(r)$. Therefore, putting 
$$A_{\rm ones} = 
A \cap \ \mbox{ones}(r), \quad A _{\rm twos} = A \cap  \ \mbox{twos}(r), \quad  A_{\rm nbubbles} = A \cap {\rm nbubbles}(r),$$
one has the disjoint union
$$A \quad = \quad A_{\rm ones} \cup A_{\rm twos} \cup A_{\rm nbubbles} \quad (A_{\rm ones} \neq A)$$
In order to impose $A \ra B$ we split $r(A  \ra B) := \{X \in r: X \ \mbox{satisfies} \ A \ra B\}$ as follows:
$$r(A\ra B)\quad =\quad \underbrace{\{X \in r: \ A \not\subseteq X\}}_{r({\tt diff})} \ \cup \ \underbrace{\{X \in r: \ A \cup B \subseteq X\}}_{r({\tt easy})} $$
From $A_{\rm ones} \neq A$ follows $r({\tt diff}) \neq \emptyset$, but $r({\tt easy}) = \emptyset$ is possible. The ``difficult'' task will be to represent $r({\tt diff})$ as a suitable disjoint union of $\{0,1,2,n\}$-valued rows.

To fix ideas, take $W = \{1,2, \cdots, 14\}$ and let $A \ra B$ be $\{1,2,3,4,5,6,7,8\} \ra \{12, 13\}$. If $r$ is the top row in Table 2 below, then the parameter $t$ in (5) is $t=4$. Furthermore 
$$A_{\rm ones} = \emptyset,\quad  A_{\rm twos} = \{1,2\}, \quad A_{\rm nbubbles} = \{3,4,5,6,7,8\}.$$
If we write $\mbox{supp}(n_1) = \{3,4,9\}$ for our first $n$-bubble, then $\mbox{supp}(n_1) \cap A_{\rm nbubbles} = \{3,4\}$.  Splitting accordingly yields
$$r({\tt diff})\quad =\quad  \underbrace{\{X \in r({\tt diff}): \ \{3,4\} \not\subseteq X\}}_{r[n]} \ \cup \ \underbrace{\{X \in r({\tt diff}): \ \{3,4\} \subseteq X\}}_{r[1]}$$
In Table 2, notice that $n_1 n_1 n_1$ in $r$ becomes $nn2$ in $r[n]$, and 110 in $r[1]$ (not shown).
Proceeding likewise with respect to $r[1]$ and $\mbox{supp} (n_2) = \{5,10\}$ we get $r[1] = r[1,n] \cup r[1,1]$, and so on. Finally $r[1,1,1] = r[1,1,1,n ] \cup r[1,1,1,1]$ where
$$\begin{array}{lll}
r[1,1,1,n] &  : = & \{X \in r[1,1,1]: 8 \not\in X\} \\
\\
r[1,1,1,1] & : = & \{X \in r[1,1,1]: 8 \in X\} \end{array}$$
It is clear that $r({\tt diff})$ is the disjoint union
$$r({\tt diff})\quad  =\quad  r[n] \cup r[1,n] \cup r[1,1,n] \cup r[1,1,1,n] \cup r[1,1,1,1]$$
As to $r({\tt easy})$, if it is nonempty like here, it can be represented as a {\it single} $\{0,1,2,n\}$-valued row.

\begin{tabular}{|c|c|c|c|c|c|c|c|c|c|c|c|c|c|l}
$1$ & $2$ & $3$ & $4$ & $5$  & $6$ & $7$ & $8$ & $9$ & $10$ & $11$ & $12$ & $13$ & $14$ & \\ \hline 
$2$ & $2$ & $n_1$ & $n_1$ & $n_2$ & $n_3$ & $n_3$ & $n_4$ & $n_1$ & $n_2$ & $n_3$ & $n_3$ & $n_4$ & $n_4$ & \quad $r$ \\ \hline \hline
$2$ & $2$ & ${\bf n}$ & ${\bf n}$ & $n_2$ & $n_3$ & $n_3$ & $n_4$ & $2$ & $n_2$ & $n_3$ & $n_3$ & $n_4$ & $n_4$ & \quad $r[n]$ \\ \hline
$2$ & $2$ & ${\bf 1}$ & ${\bf 1}$ & ${\bf 0}$ & $n_3$ & $n_3$ & $n_4$ & $0$ & $2$ & $n_3$ & $n_3$ & $n_4$ & $n_4$ & \quad $r[1,n]$ \\ \hline
$2$ & $2$ & ${\bf 1}$ & ${\bf 1}$ & ${\bf 1}$ & ${\bf n}$ & ${\bf n}$ & $n_4$ & $0$ & $0$ & $2$ & $2$ & $n_4$ & $n_4$ & \quad $r[1,1,n]$ \\ \hline
$2$ & $2$ & ${\bf 1}$ & ${\bf 1}$ & ${\bf 1}$ & ${\bf 1}$ & ${\bf 1}$ & ${\bf 0}$ & $0$ & $0$ & $n_3$ & $n_3$ & $2$ & $2$  & \quad $r [1,1,1,n]$ \\ \hline
$n$ & $n$ & ${\bf 1}$ & ${\bf 1}$ & ${\bf 1}$  & ${\bf 1}$ & ${\bf 1}$ & ${\bf 1}$ & $0$ & $0$ & $n_3$ & $n_3$ & $n_4$ & $n_4$ & \quad $r[1,1,1,1]$ \\ \hline \hline
${\bf 1}$ & ${\bf 1}$ & ${\bf 1}$ & ${\bf 1}$ & ${\bf 1}$ & ${\bf 1}$ & ${\bf 1}$ & ${\bf 1}$ & $0$ & $0$ & $0$ &${\bf 1}$ & ${\bf 1}$ & $0$ & \quad $r({\tt easy})$ \\ \hline
\end{tabular}

Table 2

Our chosen example for Case 4 was ``almost typical.'' Let us indicate the possible slight deviations: 
\begin{enumerate}
 \item [(i)] If the conclusion of $A \ra B$ was $B = \{13, 14\}$, nothing would have changed in the decomposition of $r({\tt diff})$, but then $r({\tt easy}) = \emptyset$ because $n_4n_4n_4$ cannot be $111$.
\item[(ii)] We had $A_{\rm ones} = \emptyset$, but $A_{\rm ones} \neq \emptyset$ would merely have resulted in additionally copying a bunch of $1$'s in all rows $r[n], r[1,n]$ up to $r({\tt easy})$.
\item[(iii)] Suppose $A \ra B$ was $\{3,4,5,6,7,8\} \ra \{12, 13\}$ instead. Then we have $A = A_{\rm nbubbles}$ which entails

\hspace*{1cm} $\begin{array}{lll}
r({\tt diff}) & = & r[n] \cup r[1,n] \cup r[1,1,n] \cup r[1,1,1,n]  \quad (\mbox{without} \ r[1,1,1,1]!)\\
\\
r({\tt easy}) & =& (2,2,1,1,1,1,1,1, 0,0,0, 1,1,0) \end{array}$

\item[(iv)] Suppose $A \ra B$ was $\{1,2\} \ra \{12, 13\}$ instead. Then we have $A = A_{\rm twos}$ which entails
$$\begin{array}{lll}
r({\tt diff}) & = & ({\bf n}, {\bf n}, n_1, n_1, n_2, n_3, n_3, n_4, n_1, n_2, n_3, n_3, n_4, n_4) \\
\\
r({\tt easy}) &= & ({\bf 1}, {\bf 1},  n_1, n_1, n_2, n_3, n_3, n_4, n_1, n_2, n_3, 1, 1, n_4) \end{array}$$
\end{enumerate}
It remains to show how negative clauses $A^\ast$ are imposed upon $\{0,1,2,n\}$-valued rows $r$. Matters being similar to the above we can be brief.

{\bf Case 5:} $A^\ast \cap \ \mbox{zeros}(r) \neq \emptyset$ or $A^\ast$ wholy contains a $n$-bubble. Then $r$ carries over unaltered.

{\bf Case 6:} $A^\ast \subseteq \ \mbox{ones}(r)$. Then $r$ needs to be cancelled.

{\bf Case 7:} $A^\ast \cap \ \mbox{zeros} (r) = \emptyset$ and $A^\ast$ does not wholy contain a $n$-bubble and $A^\ast \not\subseteq \ \mbox{ones}(r)$. Then with definitions analogue to case 4 one has
$$A^\ast \quad =\quad A^\ast_{\rm ones} \cup A^\ast_{\rm twos} \cup A^\ast_{\rm nbubbles} \quad (A^\ast_{\rm ones} \neq A^\ast)$$
and one treats $r({\tt diff})$ exactly as in Case 4. Note that $r({\tt easy})$  is absent in Case 7. \hfill $\blacksquare$

The encountered row splitting process is quite visual and invites hand calculations for smaller size problems.
From case 4 it is clear that for the present application of the principle of exclusion the parameter $s_{\max}$ from section 4 is at most the smaller of $\frac{w}{2}$ (since $t \leq \frac{w}{2}$ in (7)) and $\max \{|A|: \ A \ra B \ \mbox{in} \ \Sigma \}$.   
Thus it costs $O(w^2)$ to impose an implication of $\Sigma$ on $r$. Ditto by cases 5 to 7 it costs $O(w^2)$ to impose a negative clause of $\Theta$ upon $r$. Hence (13) is satisfied. Without further mention, it will be satisfied in all upcoming theorems as well.
 Recall the definition of a Horn $h$-formula from section 2.

\begin{tabular}{|l|} \hline \\
{\bf Theorem 2:} Given is a Horn $h$-formula $\Sigma \cup \Theta$ on  $w$ variables. Then  the presented Horn\\
 $n$-algorithm  can be adapted to
generate the $N$ models in time $O(hw+Nh^2w^2)$.\\
\\ \hline \end{tabular}

{\it Proof:}  The presented Horn $n$-algorithm needs to be upped from old to new according to Theorem 1 with $k =w$ (so extra feasible $=$ feasible). If we manage to satisfy (a), (b) in Theorem 1 for $f(h,w) : = hw$ (which is at least linear in $w$) then our $O(hw+Nh^2w^2) = O(f(h,w)+Nhwf(h,w))$ claim will follow.

Concerning (a), we need to show that the feasiblity of a row $r$ can be tested in time $O(f(h,w))$.
For each $Y \subseteq W$ let $\ol{Y}$ be the $\Sigma$-closure of $Y$, i.e. the smallest superset of $Y$ that satisfies all implications of $\Sigma$. As seen in Section 2, it costs $O (||\Sigma ||+w) = O(hw+w) =O(f(h,w))$ to compute $\ol{Y}$. To check the feasibility of $r$, put $Y : = ones(r)$. 
If $\ol{Y}$ is a noncover for all $A^\ast_j$ in $\Theta$ (testable in time $O(hw) = O(f(h,w))$, then $\ol{Y}$ is a $(\Sigma \cup \Theta)$-model, i.e. feasible. But is $\ol{Y} \in r$? This amounts to check, again in time $O(hw)$, whether $\ol{Y} \cap$ zeros$(r) = \emptyset$ and whether $\ol{Y}$ doesn't contain any $n$-bubble of $r$.
 It remains to show that when $X \in r$ is {\it any} $(\Sigma \cup \Theta)$-model, then so will be $\ol{Y}$. Indeed, since $X = \ol{X}$ (being a $\Sigma$-model), it follows from $Y = \ \mbox{ones}(r) \subseteq X$ that $\ol{Y} \subseteq X$. Therefore, with $X$ also $\ol{Y}$ is a noncover for all $A^\ast_j$ in $\Theta$.  
 To summarize, we have shown:
\begin{enumerate}
\item[(14)] $r$ is feasible if and only if  first $\ol{Y}$ is a noncover for all $A^\ast_j$ in $\Theta$, second $\ol{Y} \cap \ \mbox{zeros}(r) = \phi$, third $\ol{Y}$ doesn't contain an $n$-bubble of $r$. This feasibility test costs $O(f(h,w))$.
\end{enumerate}

As to (b), it is straightforward to write down all $|r|$ sets of a $\{0,1,2,n\}$-valued row $r$ in some systematic way; for instance $r = (0,1,0,n_2, n_1, n_1, 1, n_2)$ can be handled ``lexicographically'':

$\begin{array}{llll}
\{2,7\} \cup \{ \ \} \cup \{ \ \}, & \{2,7\} \cup \{ \ \} \cup \{4\}, & \{2,7\} \cup \{ \ \} \cup \{8 \} \\
\\
\{2,7\} \cup \{5 \} \cup \{ \ \}, & \{2,7 \} \cup \{5\} \cup \{4\}, & \{2,7\} \cup \{5\} \cup \{8 \} \\ 
\\
\{2,7\} \cup \{6\} \cup \{ \ \}, & \{2,7\} \cup \{6\} \cup \{4\}, & \{2,7\} \cup \{6\} \cup \{8 \}\end{array}$

Hence generating all members of a row $r$ costs $O(|r| w) = O(Card (r, \leq w) w f(h,w))$ as required.
\hfill $\blacksquare$

As to the proof of (a), if $r = (2,2, \cdots, 2)$, then $\ol{Y}$ is the closure of the empty set, and the feasbilitiy of $r$ amounts to the satisfiability of $\Sigma \cup \Theta$. 

In practise most problems are ``homogeneous'' in that either $\Sigma$ or $\Theta$ is empty. If $\Theta= \emptyset$, then only cases 1 to 4 apply and we speak of the {\it implication $n$-algorithm}. If $\Sigma = \emptyset$ then only cases 5 to 7 apply and we speak of the {\it noncover $n$-algorithm}. Its dual\footnote{Notice that $X$ is a noncover of $A^\ast_1, \cdots, A^\ast_h$ if and only if its complement $X^c = W \setminus X$ is a {\it transversal} in the sense that $X^c \cap A^\ast_i \neq \emptyset$ for all $1 \leq i \leq h$. Albeit the noncover $n$-algorithm can thus generate transversals, it pays to introduce the symbolism $ee \cdots e : =$ ``at least one 1'' and a corresponding {\it transversal $e$-algorithm} which produces the transversals ``directly'', not as $X^c$. See also the last paragraph in Section 9.4.} version is called {\it transversal $e$-algorithm} [W2]. Applications, refinements and numerical evaluations of the implication $n$-algorithm are work in progress.

{\bf 6. Generating all Horn-models of fixed cardinality}

The naive approach to $k$-element models is to retrieve (i.e. generate or count) form each final row $r$ all $X \in r$ with $|X| =k$. Trouble is $r$ may contain no such $X$ and should have been deleted long ago. Whether avoiding deletions in practise is worth the effort, depends on the sitation, but in order to get theoretic results the deletion of rows must be  ruled out. As seen, this is the task of Theorem 1. We additionally need the following fact:
\begin{enumerate}
\item[(15)] [W2, Thm.5] Let $r$ be a $\{0,1,2,n\}$-valued row. It costs time $O(w^2Card (r,k))$ to
generate, i.e. write down in set notation, the sets $X \in r$ with $|X| = k$.
\end{enumerate}
\begin{tabular}{|l|} \hline \\
{\bf Theorem 3:} Let $\Sigma \cup \Theta$ be a Horn $h$-formula on $w$ variables, and $k \leq w$ a fixed\\
integer. Then the $N$ many models with $|X| \leq k$ can be generated in time $O(hw+Nh^2w^2)$. \\ \\ \hline \end{tabular}

{\it Proof.} Again we verify (a), (b) in Theorem 1 for $f(h, w) : = hw$.  

As to (a), one checks that $r$ is extra feasible if and only if $\ol{Y} = \ol{\mbox{ones}(r)}$ satisfies the conditions stated in (14) {\it and additionally} $|\ol{Y}| \leq k$. (Notice that from $|X| \leq k$ and $\ol{Y} \subseteq X$ follows $|\ol{Y}| \leq k$). The cost stays $O(hw)$.

As to (b), by (15) it costs
$$\begin{array}{lll}
& & O(w^2 \mbox{Card}(r,1)) + O(w^2 \mbox{Card}(r,2)) + \cdots + O(w^2 \mbox{Card}(r,k)) \\
\\
& & = O(w^2\mbox{Card}(r, \leq k)) = O(\mbox{Card}(r, \leq k) w f(h,w))
\end{array}$$
to generate all $(\leq k)$-element members of $r$. \hfill $\blacksquare$

Recall that  Theorem 1 still holds when $\leq k$ is replaced by $=k$ throughout. Trouble is that in our ``Horn situation'' checking extra feasibility in condition (a) of Theorem 1 then gets more expensive. Putting it bluntly, as opposed to the proof of Theorem 3 the existence of a $(\Sigma \cup \Theta )$-model $X \in r$ with $|X| = k$ does not imply that $|\ol{Y}| = k$. 

In the present Section 6 we only tackle a special case (Theorem 4) and postpone the naked $|X| =k$ to Section 8 which is dedicated to counting, not generating. The special case is such that $h \leq w$ and $k \leq w-h$. Furthermore we focus on noncovers rather than arbitrary Horn fomulae.

\begin{tabular}{|l|} \hline \\
{\bf Theorem 4:} Given are $h$ subsets $A^\ast_j$ of a $w$-set $W$. Assume that $h \leq w$ and
fix a non-\\
negative integer $k \leq w -h$. Then the $N$ many $k$-element noncovers $X$ of the set system\\
$\{A^\ast_1, \cdots, A^\ast_h\}$ can be generated in time $O(hw+Nh^2w^2)$.\\ \\ \hline \end{tabular}

Observe that naively testing all $k$-element subsets of $W$ costs $O({w\choose k} hw) = O(w^{k+2})$ which other than $O(Nh^2w^2)$ does not involve the possibly small number $N$.

{\it Proof of Theorem 4.} We shall verify (a),(b) for $f(h,w)= hw$ in the $(=k)$-version of Theorem 1. Again (b) holds as in the proof of Theorem 3. 

It remains to show (a) i.e. that for any $\{0,1,2,n\}$-valued row $r$ its extra feasibility can be tested in time $O(f(h,w))$. 
Say the impositions of $A^\ast_{j+1}, A^\ast_{j+2}, \cdots, A^\ast_h$ upon $r$ are still pending. If one of these sets is contained in ones$(r)$, or if $|\mbox{ones}(r)| > k$, then $r$ is not extra feasible. Testing this costs $O(hw) = O(f(h,w))$. Conversely, suppose that $|\mbox{ones}(r)|\leq k$ and that $S_i : = A^\ast_i \setminus \mbox{ones}(r)$ is nonempty for all $j +1\leq i \leq h$. Looking at cases 1 to 7 in Section 5 one sees that with each imposition of a constraint the number of $n$-bubbles in a son increases by at most one (in case 4 this number even decreases in many sons). It follows that our row $r$ contains at most $j$ $n$-bubbles. Say w.l.o.g. there are exactly $j$ of them. To match previous notation, call them $S_1, \cdots, S_j$. Take any  transversal $T$ of $\{S_1, \cdots, S_h\}$ with $|T| \leq h$. A minute's reflection shows that $X: = W-T$ is a noncover of $\{A^\ast_1, \cdots, A^\ast_h\}$ and that $X \in r$.
In view of $\mbox{ones}(r) \subseteq X$ and
$$| \mbox{ones}(r)| \ \ \leq \ \ k \ \  \leq \ \ w - h \ \  \leq \ \ |X|$$
we can extend ones$(r)$ to any $k$-element subset $X_0$ of $X$. Then still $X_0 \in r$, and $X_0$ a fortiori is a noncover of $\{A^\ast_1, \cdots A^\ast_h\}$.  Hence $r$ is extra feasbile.   \hfill $\blacksquare$

{\bf 7. The principle of exclusion when aimed at counting}

Often one only needs to {\it count} rather than generate all models. Below Theorem 1 is accordingly adapted. As in the proof of Theorem 1, we let $R$ be the number of final rows produced by the POE. Admittedly the only apparent {\it theoretic} upper bound of $R$ is $N$ but we stick to $R$ to emphasize that in {\it practice} $R$ often is much smaller than $N$ (see [W2] for experiments on random problems). Like Theorem 1, also Theorem 5 holds when $\leq k$ is replaced by $\geq k$ or $=k$ throughout.

\begin{tabular}{|l|} \hline \\
{\bf Theorem 5:} Let $W$ be a $w$-set and let ${\cal P}_i \subseteq 2^W$ be constraints. Fix $k \in \{1,2, \cdots, w\}$. \\
Suppose some ``old'' version of the principle of exclusion can be employed to produce disjoint\\
 multivalued rows whose union is the set of all models. Further assume that for some\\
 function $f(h,w)$ which is at least linear in $w$, it holds that:\\
\\
(a) For each row $r$ it costs time $O(f(h,w))$ to decide whether $r$ is 
extra feasible \\
\hspace*{.5cm} in the sense of containing models $X$ with $|X| \leq k$.\\
\\
(b) If $r$ is a finalized row, then it costs $O(wf(h,w))$ to calculate $Card (r, \leq k)$.\\
\\
Then the old version can be adjusted to a new one that avoids deleting rows and \\
calculates the number $N$ of models $X \subseteq W$ with $|X| \leq k$ in time $O(f(h,w)+Rhwf(h,w))$.\\
Here $R \leq N$ is
 the number of final  rows produced by the new algorithm.\\ \\ \hline \end{tabular}

{\it Proof.} The conditions in Theorem 5 are the same as in Theorem 1, except in (b) we have $O(1 \cdot wf(h,w))$ instead of $O(Card (r, \leq k) \cdot wf(h,w))$. Since only (a) was used in Theorem 1 to establish the cost of all impositions as $O(Rhwf(h,w))$, that's also the correct corresponding cost in Theorem 5. As to the cost of counting all $(\leq k)$-element models from the final rows, by (b) in Theorem 5 it costs $O(Rwf(h,w))$ to compute the $R$ numbers Card$(r, \leq k)$. Adding up these numbers (which in base 2 have length $\leq w$) yields $N$ and costs $O(Rw)$. Hence the total cost is
$$O(Rhwf(h,w))+ O(Rwf(hw)) + O(Rw) = O(Rhwf(hw))$$ \hfill $\blacksquare$

{\bf 7.1 Space assessment}

Concerning time assessment, Theorem 5 is the twin of Theorem 1. Here we show that for the {\it counting} POE the required space can be reduced, in fact it doesn't even depend on the number $N$ of models.

Rather than calculating the often large row collections $\mbox{Mod}_i$ stepwise for $i=1$ up to $i = h$ (as we did in Section 3 for ease of visualization), it is better to employ a well known {\it  last in first out} (LIFO) stack management. That is, each row $r_j$ carries a pointer $PC(r_j)$ ($=$ pending constraint), and throughout only the {\it top} row $r_j$ of the working stack is updated. Specifically, if $PC(r_j) = k$ then constraint ${\cal P}_k$ is imposed upon $r_j$. This triggers the (trivial or proper) sons $r_{j+1}, r_{j+2}, \cdots$. They are put on the working stack in place of $r_j$, with corresponding pointers $PC$ set on $k+1$. Whenever a row is finalized, i.e. ${\cal P}_h$ has been imposed on it, it is moved from the working stack to the {\it final stack}. For instance, using LIFO the imposition of $h=4$ constraints ${\cal P}_i$ upon $r_1  = (2,2, \cdots, 2)$ may begin as follows:

\begin{tabular}{|l|} \hline
$PC(r_1) =1$ \\ \hline \end{tabular} \quad $\ra$ \quad \begin{tabular}{|l|}  \hline $PC(r_3) = 2$\\ \hline $PC(r_2) =2$ \\ \hline \end{tabular} \quad $\ra$ \quad \begin{tabular}{|l|} \hline
$PC(r_6) =3$ \\ \hline
$PC(r_5) =3$\\ \hline
$PC(r_4) =3$\\ \hline
$PC(r_2) =2$\\ \hline \end{tabular} \quad $\ra$ \quad \begin{tabular}{|l|} \hline
$PC(r_7) =4$ \\ \hline
$PC(r_5) =3$\\ \hline
$PC(r_4) =3$\\ \hline
$PC(r_2) =2$\\ \hline \end{tabular}

Fig. 1

If imposing ${\cal P}_4 = {\cal P}_h$ upon $r_7$ results in (say) the proper sons $\rho_1, \rho_2, \rho_3$, the latter are the first members of the final stack, and one proceeds by imposing ${\cal P}_3$ upon $r_5$.

\begin{center}
\includegraphics[scale=0.5]{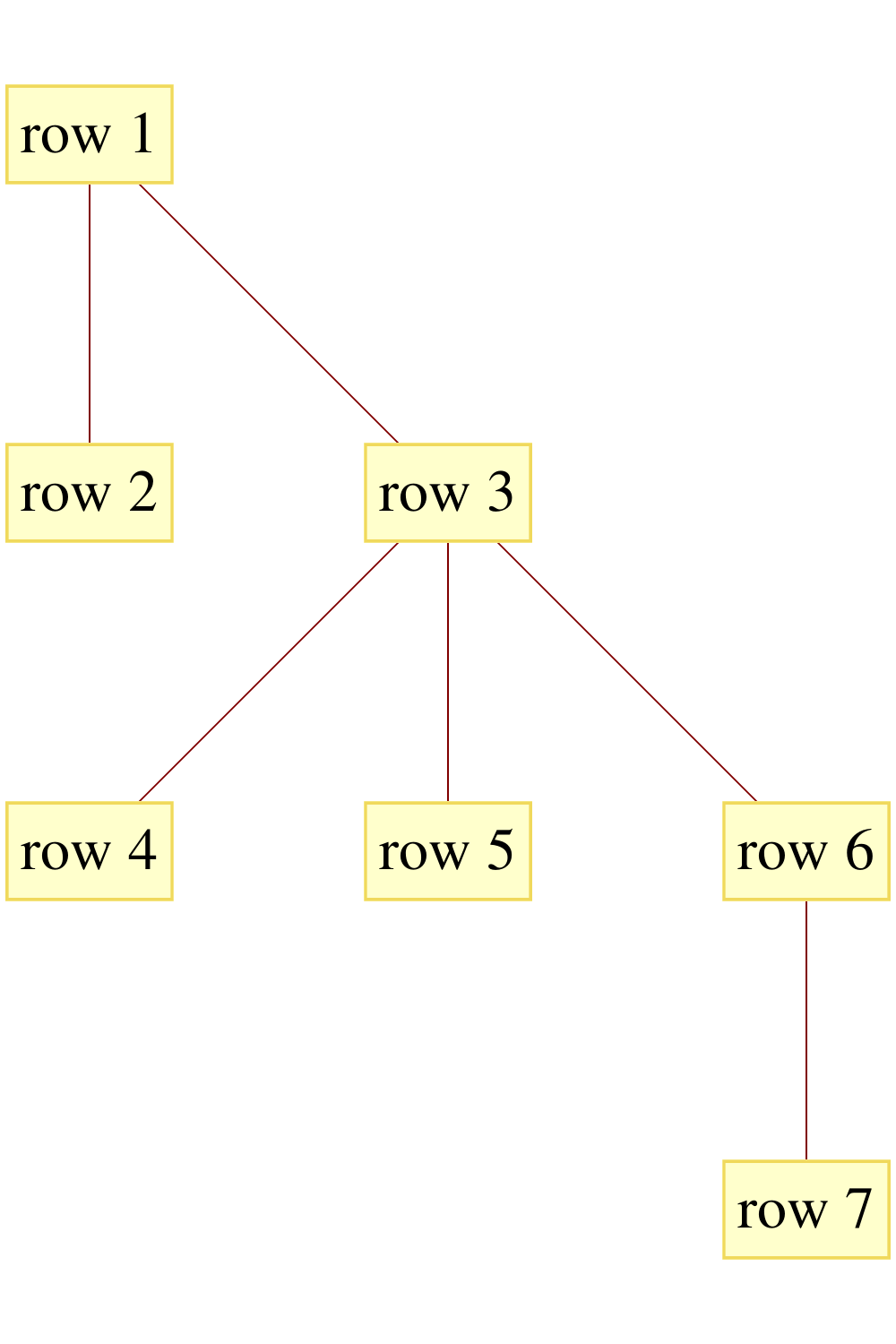}
\end{center}

The last working stack in Fig. 1 matches the rooted tree in Fig. 2 in that its four rows bijectively correspond to the leaves. It is clear that a rooted tree with maximum down-degree $s_{\max}$ and $h$ levels has at most $(h-1) (s_{\max}-1)+ h\leq hs_{\max}$ leaves. 

\begin{tabular}{|l|} \hline \\
{\bf Theorem 6:} Suppose the principle of exclusion uses LIFO to enumerate\\
(as opposed to generate) all subsets of a $w$-set that satisfy $h$ properties ${\cal P}_i$.
Let $s_{\max}$ \\
be as in Section 4. Then the whole algorithm requires space $O(s_{\max} hw)$. \\ \\ \hline \end{tabular}

{\it Proof:} As seen, using LIFO the working stack can increase to size at most $hs_{\max}$. Since each multivalued row in the working stack requires space $O(w)$, and since the final stack remains empty (final rows $r$ are thrown away after recording $|r|$), the claim follows.  \hfill $\blacksquare$

The actual row manipulations performed by the principle of exclusion (e.g. the number of deletions) are the {\it same} whether or not LIFO is used; what differs merely is the space required. Of course the LIFO-stack management is also practical for {\it generating} Horn-models (Sections 5,6), but Theorem 6 has no twin in that context.

{\bf 8. Counting all Horn-models of fixed cardinality}

We shall transfer Theorem 3 and Theorem 4 to the counting-framework as Theorem 7 and 8. Additionally two more theorems are stated. As to the repeatedly used expression ``provided $N > 0$'', see the beginning of the proof of Theorem 1.  Among the four counting-theorems in this section the first is about $(\leq k)$-element models, and the other three about $k$-element models.
Besides Theorem 5 we shall need this twin of (15):

\begin{enumerate}
\item[(16)] [W2, Thm.4] Let $r$ be a $\{0,1,2,n\}$-valued row. It costs time $O(kw^3)$ to compute the 
 $k$ numbers $Card (r,1), \cdots, Card (r,k)$.
\end{enumerate}
In Theorem 7, 8, 10 we let again $R \leq N$ be the number of final rows produced by the Horn $n$-algorithm.

\begin{tabular}{|l|} \hline \\
{\bf Theorem 7:} Given is a Horn $h$-formula $\Sigma \cup \Theta$ on $w$ variables, and a fixed integer\\
$k \leq w$. Then the $N$ many models $X$ with $|X| \leq k$ can be counted  (provided $N >0$) in time\\
$O(Rkh^2w^3) = O(Nkh^2w^3)$.\\ \\ \hline \end{tabular}

{\it Proof.} It suffices to verify (a), (b) in Theorem 5 for $f(h,w): = khw^2$, because then $N$ can be calculated in time
$$O(Rhwf(h,w)) =O(Rk^3h^2w^2)$$
In the proof of Theorem 3 we saw that the extra feasiblity of a row $r$ can be checked in time $O(hw) = O(f(h,w))$, and so condition (a) holds. As to (b), by (16) the calculation of Card$(r, \leq k) = \ \mbox{Card}(r,1) + \cdots + \ \mbox{Card}(r,k)$ costs $O(kw^3) = O(wf(h,w))$. \hfill $\blacksquare$

\begin{tabular}{|l|} \hline \\
{\bf Theorem 8:} \\
Given are $h$ subsets $A^\ast_j$ of a $w$-set $W$ such that $h \leq w$. Fix a non-negative integer\\
$k \leq w -h$. Then the number $N$ of $k$-element noncovers of the set system $\{A^\ast_1, \cdots, A^\ast_h\}$\\
can (provided $N > 0$) be calculated in time $O(Rkh^2w^3) = O(Nw^6)$. \\ \\ \hline \end{tabular}

{\it Proof.} If we manage to verify (a), (b) for $f(h,w) : = khw^2$, in the $(=k)$-version of Theorem 5, the $O(Rhwf(h,w))= O(Rkh^2w^3)$ claim will again follow. We have (a) because of $O(hw) = O(f(h,w))$. As to (b), by (16) calcualting Card$(r,k)$ costs $O(kw^3) = O(wf(h,w))$; unfortunately Card$(r,k)$ isn't cheaper than Card$(r, \leq k)$ before. \hfill $\blacksquare$

A natural idea is to calculate the number $N$ of models $X$ with $|X| =k$ as $N = N'-N''$, where $N'$ and $N''$ are the easier numbers of models of cardinality $\leq k$ and $\leq (k-1)$ respectively. Unfortunately, albeit unlikely in practise, $N'$ and $N''$ may grow exponentially with respect to $N$. Nevertheless, the idea can be saved in this form:

\begin{tabular}{|l|} \hline \\
{\bf Theorem 9:} Given is a Horn $h$-formula on $w$ variables and an integer $k \leq w$.\\
 Suppose it is known that the number of $k^\ast$-element models increases as $k^\ast$ increases\\
 from $1$ to $k$. Then the $N$ models $X$ with $|X| =k$ can (provided $N > 0$) be counted in time\\
 $O(Nk^2h^2w^32)=O(Nh^2w^5)$. \\ \\ \hline  \end{tabular}

{\it Proof.} Let $N'$ and $N''$ be as above. Because of our assumption about increasing $k^\ast$-levels, we have $N' \leq Nk$ and $N'' \leq N(k-1)$. From $N = N' - N''$ and Theorem 7 hence follows that calculating $N$ costs $O(N'kh^2w^3) + O(N''kh^2w^3) = O((Nk)kh^2w^3)$. \hfill $\blacksquare$

Our last theorem confronts the naked $|X|=k$ condition.
To prepare for it, consider this $\{0,1,2,n\}$-valued row $r$ of length thirteen:

\hspace*{2cm} \begin{tabular}{l|c|c|c|c|c|c|c|c|c|c|c|c|c|} \hline 
& 1 & 2 & 3 & 4&5 &6 & 7 & 8 & 9 & 10 & 11 & 12 & 13\\ \hline
$r=$ & $0$ & $n_1$ & $n_1$ & $n_1$ & $n_2$ & $n_2$ & $1$ &  $2$ & $n_3$ & $n_3$ & $n_3$ & $n_3$ & 1\\ \hline
$r_0=$ & ${\bf 0}$ & $n$ & $n$ & ${\bf 1}$ & ${\bf 1}$ & $0$ & ${\bf 1}$ & ${\bf 1}$ & ${\bf 0}$ & $2$ & $2$ & ${\bf 1}$ & $1$ \\ \hline \end{tabular}

Let's calculate the number $N_0$ of $X \in r$ that satisfy neither of the implications $\{4,5\} \ra \{9\}$ and $\{5,7,8\} \ra \{1\}$, nor the negative clause $\{4,8,12\}^\ast$, and that have cardinality $|X| \neq 8$. The failure of the three formulae forces the boldface entries in row $r_0$; they in 
turn trigger all the further differences between $r_0$ and $r$. Since $|\mbox{ones}(r_0)| = 6$ the number of $X \in r_0$ with $|X| = 8$ is the number of ways to place exactly two $1$'s upon $nn22$. Ad hoc this number evaluates to $5$, and so $N_0 = |r_0| -5 = 7$. Notice that the argument above necessitates {\it unit} implications, i.e. having singleton conclusions.

\begin{tabular}{|l|} \hline \\
{\bf Theorem 10:} Given is a Horn $h$-formula $\Sigma \cup \Theta$ on $w$ variables such that $\Sigma$ consists\\
 of unit implications.   For any  integer $k \leq w$ the $N$ many models $X$ with $|X| =k$ \\
can (provided $N > 0$) be counted in time $O(R2^hhw^4k)= O(N2^hhw^5)$. \\ \\ \hline \end{tabular}

Since each  implication $A \ra B$ can be split into $|B|$ unit implications, Theorem 10 really handles arbitrary sets of Horn formulae. The factor $2^h$ doesn't look so ugly if one recalls that naively checking all $k$-element sets costs $O(w^{k+2})$. For instance, letting $h= \alpha w$ ($\alpha \geq 0$ fixed) and $k = \frac{w}{\beta} \ (\beta \geq 1$ fixed) one has $2^h/w^{k+2} \ra 0$ as $w \ra \infty$.

{\it Proof of Theorem 10.} We put $f(h,w) = k2^hw^3$ and verify (a), (b) in the $(=k)$-version of Theorem 5. The claim then follows from $O(R hwf(h,w)) = O(R2^hhw^4k)$. As to (b), it is satisfied since by (16) calculating Card$(r,k)$ costs $O(kw^3) = O(wf(h,w))$.

As to (a), let $a_j$ be the property of any $Y \in r$ that it satisfies the $j$-th component formula in $\Sigma \cup \Theta$ $(1 \leq j \leq h)$. Further let $a_{h+1}$ be the property that $|Y| =k$. If $N(a_1, \cdots, a_{h+1})$ is the number of $Y \in r$ satisfying all properties, then $r$ is extra feasible if and only if $N(a_1, \cdots, a_{h+1}) > 0$. The latter number by inclusion-exclusion is
$$\begin{array}{lll} N(a_1, \cdots, a_{h+1}) & = & |r| - N(\ol{a}_1) - \cdots - N(\ol{a}_{h+1}) + N(\ol{a}_1, \ol{a}_2) + \cdots + N(\ol{a}_h, \ol{a}_{h+1})\\
\\
& &  - N(\ol{a}_1, \ol{a}_2, \ol{a}_3) - \cdots \pm N(\ol{a}_1, \ol{a}_2, \cdots, \ol{a}_{h+1}) \end{array}$$
where e.g. $N(\ol{a}_3, \ol{a}_{h+1})$ denotes the number of $Y \in r$ that {\it violate} properties $a_3$ and $a_{h+1}$. As seen in the example, each summand $N_0$ involving $\ol{a}_{h+1}$ can be calculated as $N_0=|r_0| - Card (r_0,k)$ for some immediately derived row $r_0$. If $\ol{a}_{h+1}$ is not occuring, the calculation boils down to $N_0 =|r_0|$.  By (16) computing $Card (r_0, k)$ costs $O(kw^3)$, and so calculating $N(a_1, \cdots, a_{h+1})$ costs $O(2^hkw^3)= O(f(h,w))$.  \hfill $\blacksquare$

Using again $f(h,w) = k2^hw^3$ but Theorem 1 instead of Theorem 5 one immediately derives that $O(Nhwf(h,w)) = O(N2^hhw^5)$ is the cost when ``counting'' is substituted by ``generating'' in Theorem 10.

\vskip 1cm

{\bf 9 Positioning the principle of exclusion}

As seen, the POE is concerned with the models of Boolean functions $b: \{0,1\}^w \ra \{0,1\}$ when $b$ is given as a conjunction of $h$ suitable subformulae. Usually we employ set theoretic terminology and thus speak of $h$ constraints ${\cal P}_i \subseteq 2^W$. More about ``suitable'' in Section 9.4. The forthcoming comparison of POE with other methods is  preliminary and is cut along these basic tasks concerning NP-hard problems: 
\begin{itemize}
 \item Count all models or all  $k$-element models
\item Generate all models
\item Find one best (or all best) models with respect to a target function $f(x)$
\end{itemize}

{\bf 9.1 Counting all models or all $k$-element models}

The POE struggles to compete with binary decision diagrams (BDD),  now incorporated in Mathematica 8.0, when it comes to counting {\it all} 
models  of the Boolean function $b: \{0,1\}^w \ra \{0,1\}$. In brief, the BDD associated to $b(x)$ is a certain directed graph with among other nodes a root of indegree zero and two endnodes {\tt True} and {\tt False} of outdegree zero. Except for the endnodes all nodes have exactly two outgoing arcs labelled 0 and 1. Any bitstring $x \in \{0,1\}^w$ triggers an obvious directed path that starts at the root and ends at either {\tt True} or {\tt False}, depending on whether $b(x) =1$ or $b(x) =0$. Starting at the bottom of the BDD, it is well known and easy  to recursively compute the {\it exact} probability $p$ that a {\it random} bitstring $x \in \{0,1\}^w$ triggers a path that ends at {\tt True}. It follows that $|\mbox{Mod}_h|$ can be calculated with lightning speed as $p2^w$. Faster still is it to decide whether or not some Boolean function is merely satisfiable. (Something good comes out of that for POE. Namely, a multivalued row $r$ often readily spawns a Boolean function $b_r(x)$ such that the feasibility of $r$ amounts to the satisfiability of $b_r(x)$. This needs further exploration.) 

The BDD approach nevertheless looses out on POE regarding counting models of {\it fixed cardinality} $k$ (concerning the relevance of this task, see [BEHM]).  The reason is that e.g. setting up in DNF a Boolean function $\beta : \{0,1\}^{25} \ra \{0,1\}$ which is {\tt True} exactly on the bitstrings of weight $k = 12$ already causes a memory complaint. Let alone building a BDD for the desired compound Boolean function $b(x) \wedge \beta (x)$; more details in [W2].

{\bf 9.2 Generating all models}

All models $X \in \ \mbox{Mod}_h$ can sometimes be generated one by one with ``polynomial delay'', e.g.  by so called combinatorial Gray-codes [S]. Closer to the compact encoding of $\mbox{Mod}_h$ achieved by the POE, is the BDD and the $0,1$ integer programming (01IP) framework. Indeed, both are fit to yield $\mbox{Mod}_h$ in LIFO fashion (thus no space problem) as a disjoint union of $\{0,1,2\}$-valued rows. But the POE, due to its use of {\it additional symbols} (say $n$), is more flexible and hence tends to produce much fewer multivalued rows. Ditto deletions of branches in the 01IP search tree are more frequent than in the POE search tree. Interestingly, in the case of a BDD the (albeit many) $\{0,1,2\}$-valued rows can be obtained without deletions [A, p.22]. Different from Theorem 1 and 5 this doesn't yield theoretic results since the cost of constructing the BDD itself is usually impossible to assess; a notable exception is [Be].

Returning to the search trees of 01IP and POE, their main difference is not the frequency of cutting branches, nor the fact that one is binary and the other (usually) not, but the cause of branching in the first place. While 01IP-branching is due to setting {\it variables} $x_i$ to $0$ or $1$, the POE-branching is triggered by imposing new {\it constraints} upon multivalued rows. In this regard the POE bonds more with constraint programming (CP) than with BDD or 01IP. While the POE so far has all variables $x_i$ in $\{0,1\}$, in constraint satisfaction problems [FA, ch.12] the $x_i$'s can assume values from larger but finite domains $D_i \ (1 \leq i \leq w)$. However, there is no reason to stick with $\{0,1\}$ in future application of POE. A more telling difference between CP and POE is this: Upon applying so called {\it constraint propagation} the domains shrink to $D'_i \subseteq D_i$ (if some $D_i = \emptyset$, the problem is infeasible). Different from ``POE constraint propagation'' which concisely yields $\mbox{Mod}_h$, CP constraint propagation only delivers $\mbox{Mod}_h$ as an (using universal algebra talk) unknown subdirect product of $D_1'  \times \cdots \times D_w'$.

{\bf 9.3 Finding one best or all best models}

If only one best model (w.r. to $f(x)$) needs to be found, then all of $\mbox{Mod}_h$ needs to be filtered anyway, and so the subdirect product complaint evaporates. Besides explicite enumeration and checking of models (say naively or with combinatorial Gray codes), the only known techniques to solve a NP-hard optimization problem {\it exactly} are branch and bound or dynamic programming. The POE in optimization mode falls under the branch and bound hat, along with 01IP and CP. What we called (weakly) feasible is an essential notion in any branch and bound algorithm.  It is illustrative to contrast our weak feasibility with the CP concept of {\it arc consistency} [FA]. Both concepts refer to an individual constraint, but CP is about existence of variables in $D'_i$ while POE is about a whole multivalued row.

When it comes to the {\it approximate} solution of NP-hard optimization problems, other techniques enter the stage: Simulated annealing, tabu search, genetic algorithms. Genetic algorithms bear a vague resemblance to POE in that a whole ``population'' of models is kept alive, but otherwise they differ a lot from POE. Of course also branch and bound can be switched to approximation mode: If $f(x)$ needs to be maximized, choose $t \in \R$ and apply branch and bound to either find some $x_0$ with $f(x_0) \geq t$, or to conclude that there are no such $x_0$. It has been pointed out that POE branch and bound easily yields {\it all} $x_0$ with $f(x_0) \geq t$, whereas 01IP-branch and bound is hard pressed to do the same.

{\bf 9.4 Two final remarks}

First, the POE can also be applied to disjunctions as opposed to conjunctions of properties. Specifically, in order to find the {\it cardinality} $N$ (not the members themselves) of the set ${\cal P}_1 \cup \cdots \cup {\cal P}_h$, apply the standard POE to the negated constraints ${\cal P}_i^c$ and get $N = 2^w - |{\cal P}_1^c \cap \cdots \cap {\cal P}_h^c|$. For instance, the cardinality of a  simplicial complex can thus be determined from its facets.

Second, what ``suitable subformulae'' meant at the beginning of Section 9, is merely that they allow a POE implementation that is efficient in practise, whether or not that can be backed theoretically. For instance in [W1] the suitable subformulae nicely match the stars of a graph (all edges incident with a vertex form a star) but a  theoretic assessment of the resulting swift algorithm eluded the author. If the suitable subformulae are negative clauses or implications, then theory (present article) and practise (e.g. [W2]) go hand in hand. 

Recall that {\it every} Boolean function is equivalent to a conjunction of clauses, but different from Section 2 such a clause can have more than one positive literal; say $\ol{a}_1 \vee \ol{a}_2 \vee a_3 \vee a_4 \vee a_5$. Nevertheless, being equivalent to $(a_1 \wedge a_2) \ra (a_3 \vee a_4 \vee a_5)$, we may view it as an $(\wedge, \vee)$-{\it implication}, and observe that its models are contained in the disjoint multivalued rows $(n,n,2,2,2)$ and $(1,1,e,e,e)$. (The dual $e$-symbolism was explained in Section 6.) Simultaneous occurence of $n,e$ complicates a theoretic treatment, yet the resulting POE implementation performs well in practise (work in progress).

{\bf Acknowledgement:} I am grateful to Egon Balas for insightful comments, and to an anonymous referee for a quality of constructive criticism that I haven't experienced in a long time.

{\bf References}
\begin{enumerate}
\item[{[A]}] H.R. Andersen, An introductionn to Binary Decisiion Diagrams, Lecture Notes IT University Copenhagen 1999.
\item[{[Be]}] M. Behle, On threshold BDD's and the optimal variable ordering, J. Comb. Optim. 16 (2008) 107-118.
\item[{[BEHM]}] M. Bruglieri, M. Ehrgott, H.W. Hamacher, F. Maffioli, An annotated bibliography of combinatorial optimization problems with fixed cardinality constraints, Disc. Appl. Math. 154 (2006) 1344-1357.
\item[{[BM]}] K. Bertet, B. Monjardet, The multiple facets of the canoncial direct unit implicational base, Theoretical Computer Science 411 (2010) 2155-2166. 
\item[{[FA]}] T. Fr\"{u}hwirth, S. Abdennadher, Essentials of constraint programming, Springer Verlag 2003.
\item[{[GW]}] B. Ganter, R. Wille, Formal Concept Analysis. Mathematical Foundations, Springer Verlag 1999, 284 pages.
\item[{[KOV]}] P. Krajca, J. Outrata, V. Vychodil, Parallel recursive algorithm for FCA, Proc. CLA 2008, pp.71-82.
\item[{[MR]}]H. Mannila, K.J. R\"{a}ih\"{a}, The design of relational databases, Addison-Wesley 1992. 
\item[{[S]}] C. Savage, A survey of combinatorial Gray codes, SIAM Rev. 39 (1997) 605-629.
\item[{[SS]}] U. Sch\"{o}ning, R. Schuler, Renamable Horn-clauses and unit resolution, Tech. Report, University Koblenz 1989.
\item[{[W1]}] M. Wild, Generating all cycles, chordless cycles, and Hamiltonian cycles with the principle of exclusion, Journal of Discrete Algorithms 6 (2008) 93-102.
\item[{[W2]}] M. Wild, Counting or producing  all fixed cardinality transversals, submitted, available in the arxiv. 
\item[{[W3]}] M. Wild, Computing the output distribution of a stack filter from the DNF of its positive Boolean function, to appear in Information Processing Letters.
\end{enumerate}
\end{document}